\documentclass[runningheads]{svmult}

\usepackage{makeidx}   
\usepackage{graphicx}  
\usepackage{subeqnar}  
\usepackage{multicol}  
\usepackage{physprbb}  
\makeindex             

\begin{document}
\title*{Obscuration and circumnuclear medium in nearby and distant AGN}
\toctitle{Obscuration in nearby and distant AGN}
%
%
\titlerunning{Obscuration in nearby and distant AGN}
%
\author{R. Maiolino}
\authorrunning{R. Maiolino}
%
%
\institute{INAF - Osservatorio Astrofisico di Arcetri,
Firenze, Italy}

\maketitle              

\begin{abstract}
Some recent results on the physical and statistical properties of nearby and distant
AGN are presented. I first discuss
the properties of ``elusive'' AGNs,
i.e. obscured AGNs which do not show a Seyfert-like spectrum in the optical. Then
I present preliminary results from a detailed study of
the contribution of obscured AGNs and of their host galaxies to the infrared
cosmic background. Finally I discuss an observational program aimed
at investigating the properties of the most distant quasars, of their
circumnuclear medium and the implications for their host galaxies.
\end{abstract}

\section{Introduction}

Obscured Active Galactic Nuclei (type 2 AGNs) have been investigated in great
detail in the local universe. 
The recent deep surveys have also found several
obscured AGN at high and intermediate redshift, and
have shed new light on their evolution \cite{hasinger03}.
However, there are still some open issues on the physical and statistical
properties of obscured AGNs. There is growing evidence for a population of obscured
AGNs which do not show the classical AGN signatures in their optical
spectra, which might require a revision of the unified model
as well as reassessment of the density of AGNs in the local universe.
At higher redshift it is now clear that a mixture of obscured and unobscured AGN produce
most of the X-ray background, but it is much less clear what is their
contribution to the IR background and the possible connection with
the evolution of galaxies. At the
highest redshift (z$\sim$6) probed so far by quasar surveys, it
is not clear what are the properties of the circumnuclear medium and,
specifically, if the gas has the requirements to produce obscuration,
both in terms of metallicity 
and dust content.
In this paper I shortly summarize some recent work aimed at tackling
these issues.

\section{Elusive AGNs}
A fraction of active galactic nuclei do not show the classical
Seyfert-type signatures in their optical spectra, i.e. they are optically
``elusive''.The closest example of this class of objects is NGC4945.
This galaxy hosts a nuclear starburst and its optical spectrum is
characterized by faint LINER-like emission lines associated with
the starburst superwind. However, its hard X-ray
spectrum has revealed the presence of a heavily obscured AGN \cite{guainazzi00}.
Another clear case has been reported by \cite{dellaceca02}, who detected
a heavily obscured AGN in the starburst/HII system NGC3690.
We specifically define ``elusive AGN'' as those
nuclei which do not show Seyfert-like emission lines in their optical
spectra, but where a relatively luminous AGN (i.e. in the
Seyfert range) is detected at other wavelengths. Although this class
of AGNs clearly exist, it is not clear how common they are, nor
it is clear their nature (i.e. why they are optically elusive).

\begin{figure}[b]
\begin{center}
\includegraphics[width=.47\textwidth]{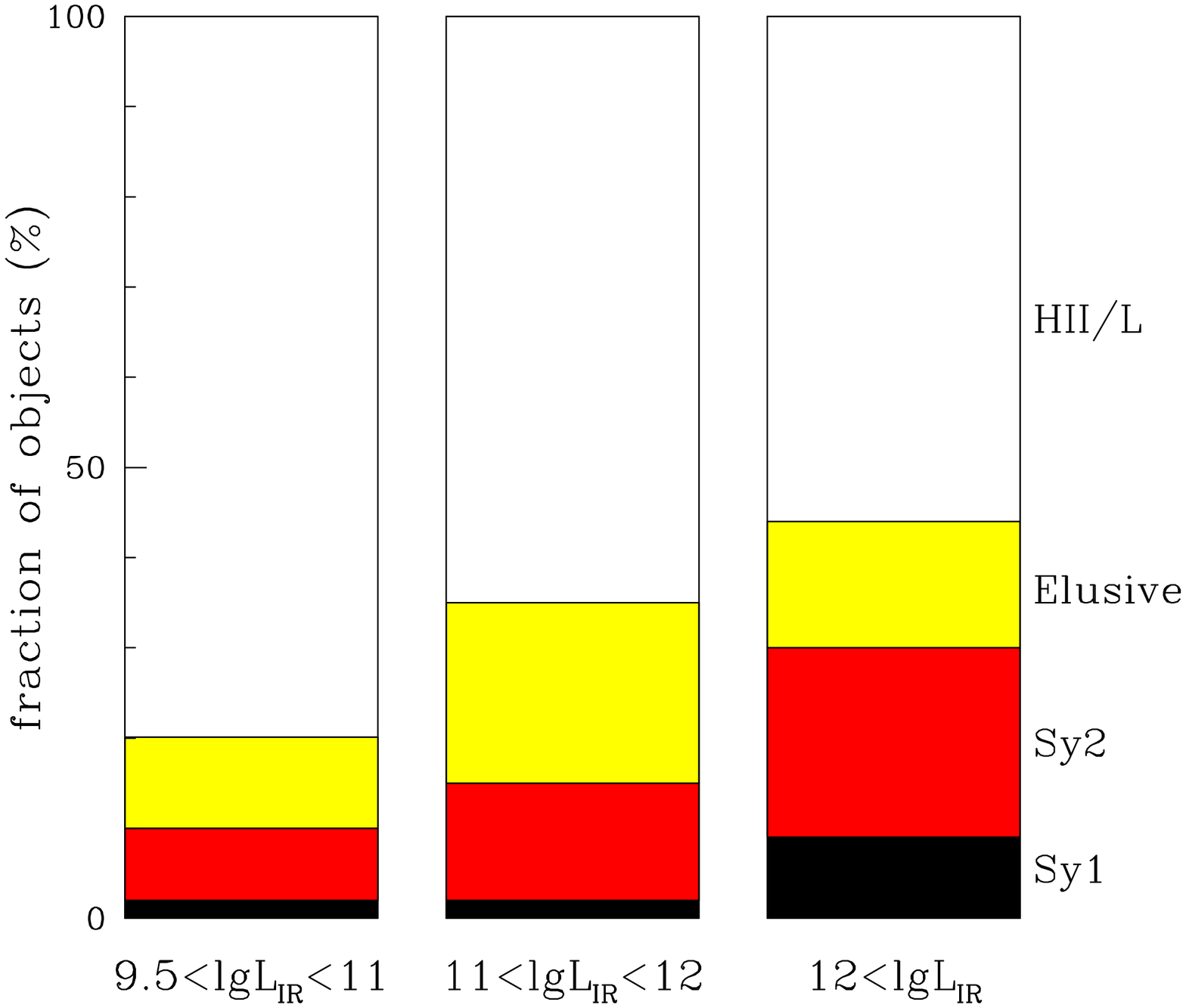}
\hspace{0.2cm}
\includegraphics[width=.47\textwidth]{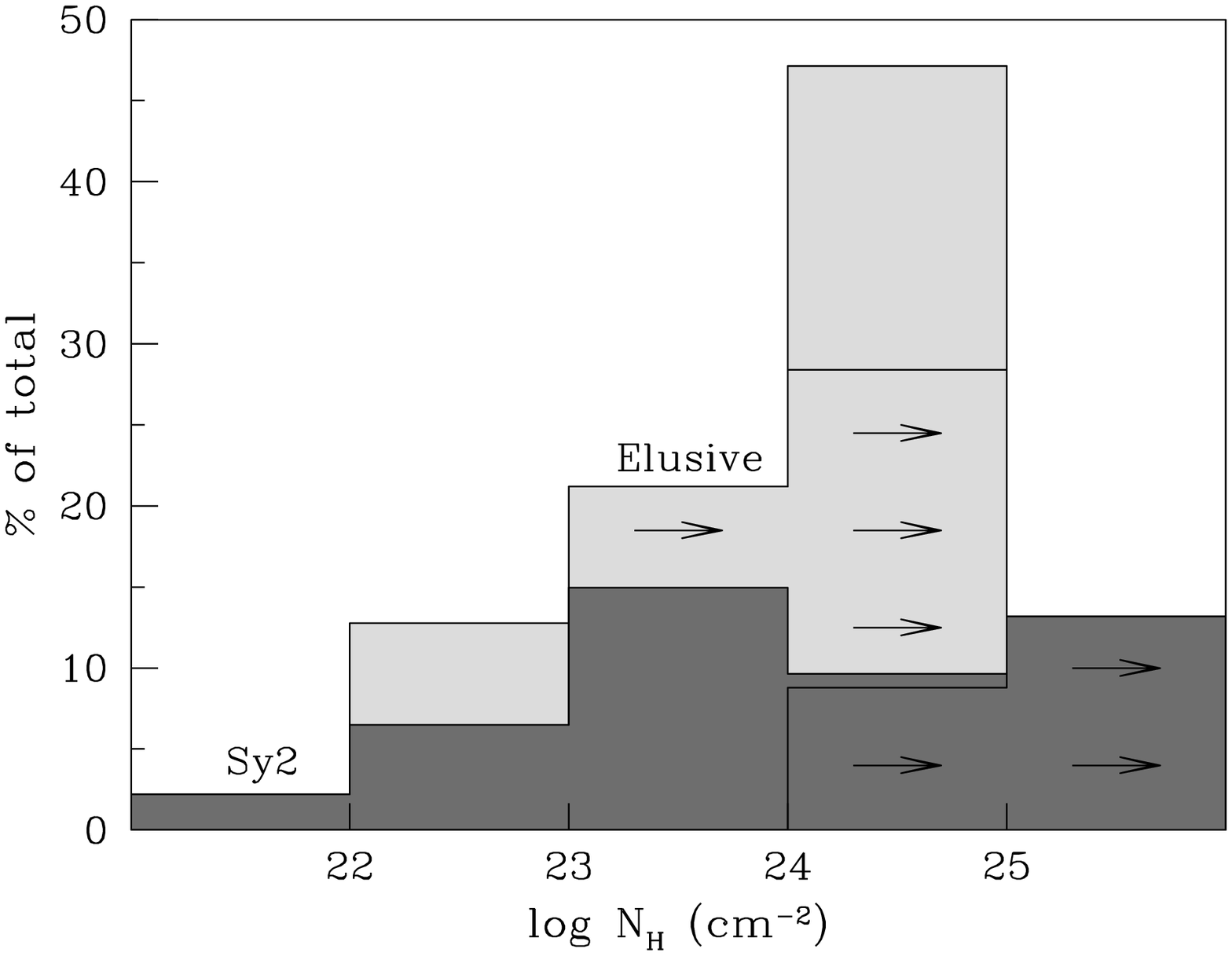}
\end{center}
\caption[]{{\it Left.} Fraction of elusive AGN and Seyfert nuclei as
a function of IR luminosity. {\it Right.} Cumulative absorbing $\rm N_H$
distribution including elusive AGNs. From \cite{maiolino03a}}
\end{figure}

We have started a program aimed at assessing
the fraction of elusive AGN in the local universe and to investigate
their nature. We are mostly exploiting hard X-ray data from Chandra
and XMM, but also near- and mid-IR spectroscopy, to detect
obscured AGNs not identified by optical spectroscopy. Preliminary
results were reported in \cite{maiolino03a} and summarized here.
There are about 20 elusive AGN identified
so far (though not all of them can be used for statistical purposes,
see \cite{maiolino03a}). Once selection
effects are taken into account we estimate that elusive AGNs may
be as numerous as (or even outnumber) classical, optically identified
Seyfert nuclei. The estimated fraction of elusive AGNs as a function of
infrared luminosity is shown in Fig.1. Obviously the statistics
are still poor and more data are required to secure this
result. If confirmed, an important implication would be that
the overall fraction of galaxies
hosting an AGN in the local universe is significantly higher than
estimated previously by optical surveys. This would nicely match the
recent results from the hard X-ray surveys
\cite{ueda03}\cite{hasinger03}\cite{fiore03},
which are finding that the evolution of
Seyfert nuclei peaks at much lower redshifts
than quasars and probably requiring a high density of Seyferts at z$=$0.
Note that the higher redshift counterparts of elusive AGN
may be the so-called XBONGs (X-ray Bright Optically Normal Galaxies)
found in the hard X-ray surveys \cite{comastri02}.

A most interesting result of the X-ray spectra of elusive AGNs
is that they are heavily absorbed and, in particular, most of them
are Compton thick, i.e. absorbed by column
of gas $\rm N_H > 10^{24}~cm^{-2}$. This suggests that their elusive
nature is associated with heavy obscuration.
In \cite{maiolino03a} we suggested that probably in elusive AGNs the 
nuclear radiation source is
obscured in all directions thus preventing
UV photons to escape and to produce a Narrow Line Region (NLR). However,
for some of the elusive AGNs other explanations
are viable, such as heavy obscuration of
the NLR and/or dilution by the circumnuclear
starburst (which may apply to the more distant cases)\cite{concalves99}.
An important implication of the heavy obscuration affecting elusive
AGNs is that, if they are included in the local census of AGNs,
then the overall distribution of absorbing $\rm N_H$ would be
strongly skewed towards higher values with respect to what
previously estimated for optically identified Seyfert 2,
as shown in Fig.1.

Another interesting feature is
that those elusive AGNs which were observed also in the 10--100~keV band
result to be absorbed by columns in the range $\rm 10^{24}<N_H<10^{25}~cm^{-2}$.
In this case the observed X-ray spectrum is characterized 
by a prominent bump peaking at about 30--40~keV. If elusive
AGNs are common also at higher redshift, then they could contribute significantly
to the 30~keV bump of the X-ray background. Within this context, one should keep in mind that
the X-ray background has been resolved at energies $<$10~keV, but the bulk
of the X-ray background, which is produced at $>$10~keV, has not been
resolved yet.

\section{The contribution of AGNs to the infrared background}

Although it is clear that most of the X-ray background is produced
by a mixture of obscured and unobscured AGN (the latter dominating
at higher energies), it is less clear what is their
contribution to the infrared cosmic background. This is a most important
issue, since the infrared is the spectral region were most of the
cosmic background is produced (after the CMB) and which is expected
to trace a significant fraction of the global star formation history.
Dust in the circumnuclear region of AGN absorbs most
of the optical-UV light emitted by the nucleus and which is reprocessed into
the infrared; therefore the contribution of AGN (and in
particular {\it obscured} AGNs) to the IR background
may be significant. Some previous studies have obtained contradictory
results depending on the assumed Spectral Energy Distribution (SED) and
evolution of the AGNs.

\begin{figure}[h]
\begin{center}
\includegraphics[angle=-90,width=\textwidth]{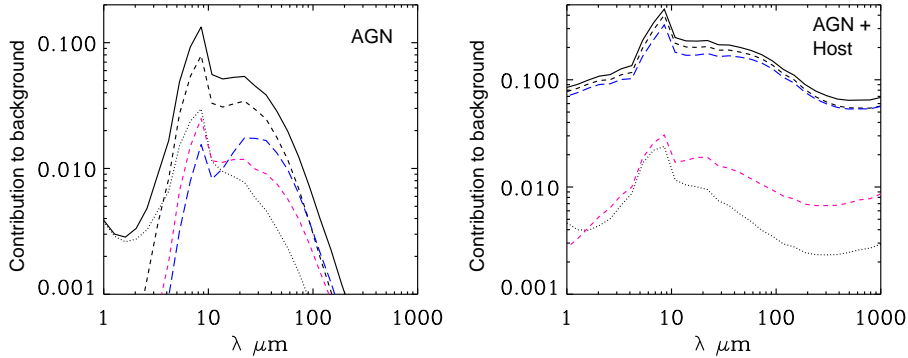}
\end{center}
\caption[]{{\it Left.} Fractional contribution of AGN to the IR background.
Dotted and dashed lines show the relative contribution by AGN affected by
different degrees of obscuration, while the solid line is the total
contribution. {\it Right.} The contribution to the IR background when AGN
host galaxies are also included.}
\end{figure}

We have approached the problem by minimizing
the number of assumptions and by using known and measured quantities 
to bridge the (AGN-produced) X-ray background to the IR background.
In particular, we used the AGN hard X-ray luminosity functions and
evolution recently derived by \cite{ueda03} using a compilation
of the most recent X-ray surveys. We then derived the {\it nuclear} infrared
SED of a sample of about 30 Seyferts by using high resolution near-
and mid-IR data. Such IR SEDs were then normalized to the intrinsic
hard X-ray emission. Then we divided the
SEDs in bins of absorbing $\rm N_H$; this novel approach allows to
consistently use the same method adopted for the synthesis modelling of
the X-ray background. With this information we can estimate the AGN contribution
to the IR background starting from the X-ray background
and by essentially using only observed quantities.

Further details will be provided in a forthcoming paper \cite{silva03},
here we only discuss some preliminary results. Fig.2 shows the estimated
fractional contribution of AGNs to the cosmic background at various
infrared wavelengths. The contribution is low at all wavelengths: the maximum
contribution (but still $\le$10\%) is in the mid-IR, between 5$\mu$m and 40$\mu$m;
while the contribution is negligible at longer wavelengths.
However, AGNs are hosted in galaxies which may contribute significantly to
the global infrared light emitted by these systems. If the host
galaxies co-evolve with the AGN (and there are observational indications
in favor of this scenario), then they could contribute significantly
to the IR background. We have investigated this possibility by
associating the SED of the host galaxies to the nuclear SEDs. We have
then assumed that the host galaxies are linked to the AGN by the same
cosmic evolution. The resulting contribution to the IR background is shown
in Fig.2. AGNs and their host galaxies appear to contribute significantly
to the IR background: about 20\% in the mid-IR, exceeding 40\% at 8$\mu$m,
and $\sim$10\% in the far-IR and sub-mm. Therefore, the cosmic IR background
may have the imprint of the co-evolution of AGN and galaxies. Our results
are in good agreement with the contribution at 15$\mu$m obtained by
\cite{fadda02}\cite{alexander02} by cross-correlating
ISO and Chandra/XMM sources. Additional implications, especially for what
concerns to contribution to the IR/submm source number counts will be
discussed in \cite{silva03}.

\section{Absorption and circumnuclear medium in the most distant quasars}

The most distant quasars known are at a redshift (z$\sim$6) which is approaching
the epoch of reionization. At early epochs it is not obvious that
the circumnuclear gas has the requirements to obscure AGNs. Indeed,
obscuration in the X-rays requires that the gas metallicity has already
evolved enough to provide the opacity associated to various elements,
and in particular iron (whose edge is shifted into the soft X-rays).
Optical and UV absorption require that dust has already been produced
in large quantities at such high redshifts.


\begin{figure}[h]
\includegraphics[angle=0,width=.40\textwidth]{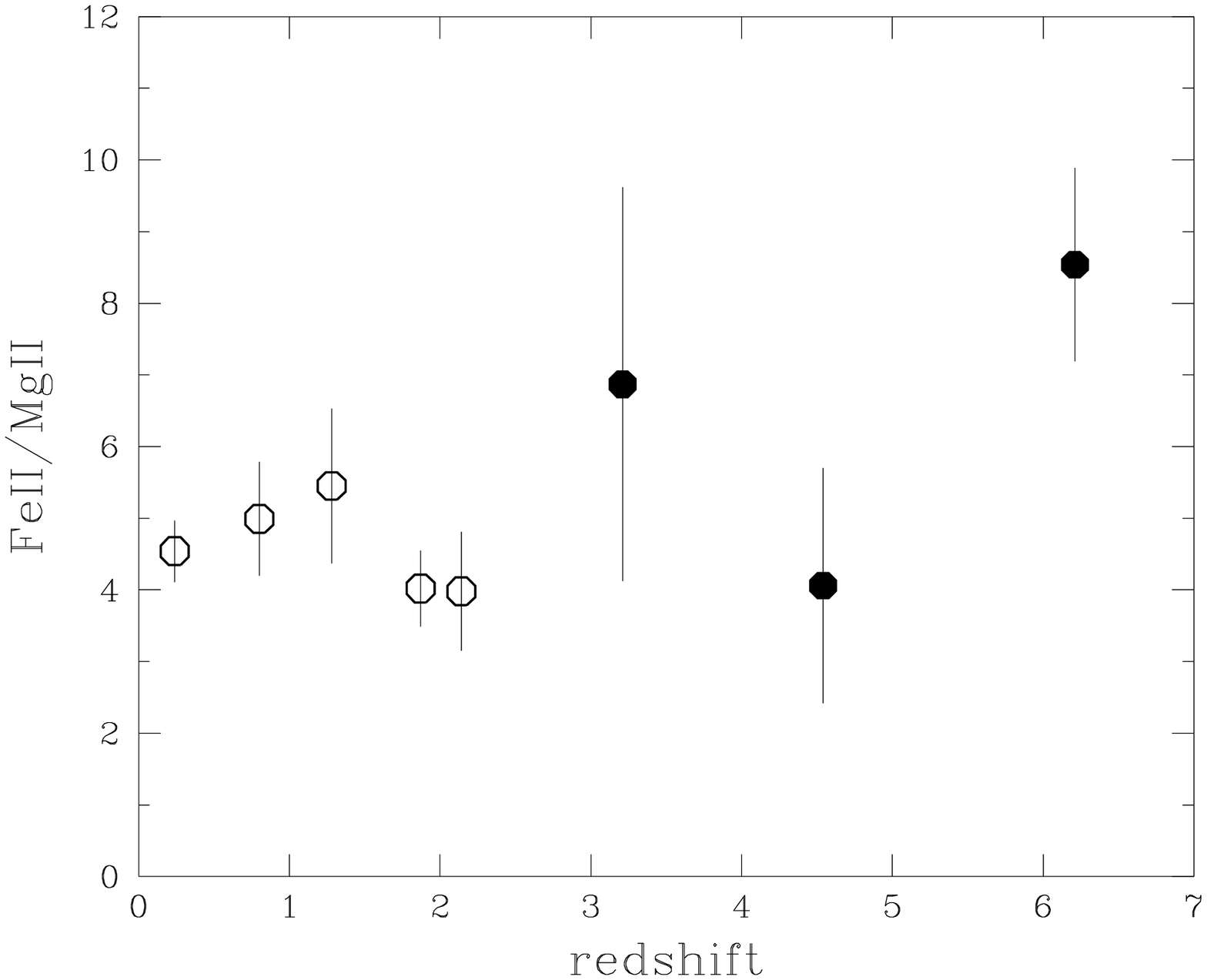}
\hspace{0.1cm}
\includegraphics[angle=0,width=.55\textwidth]{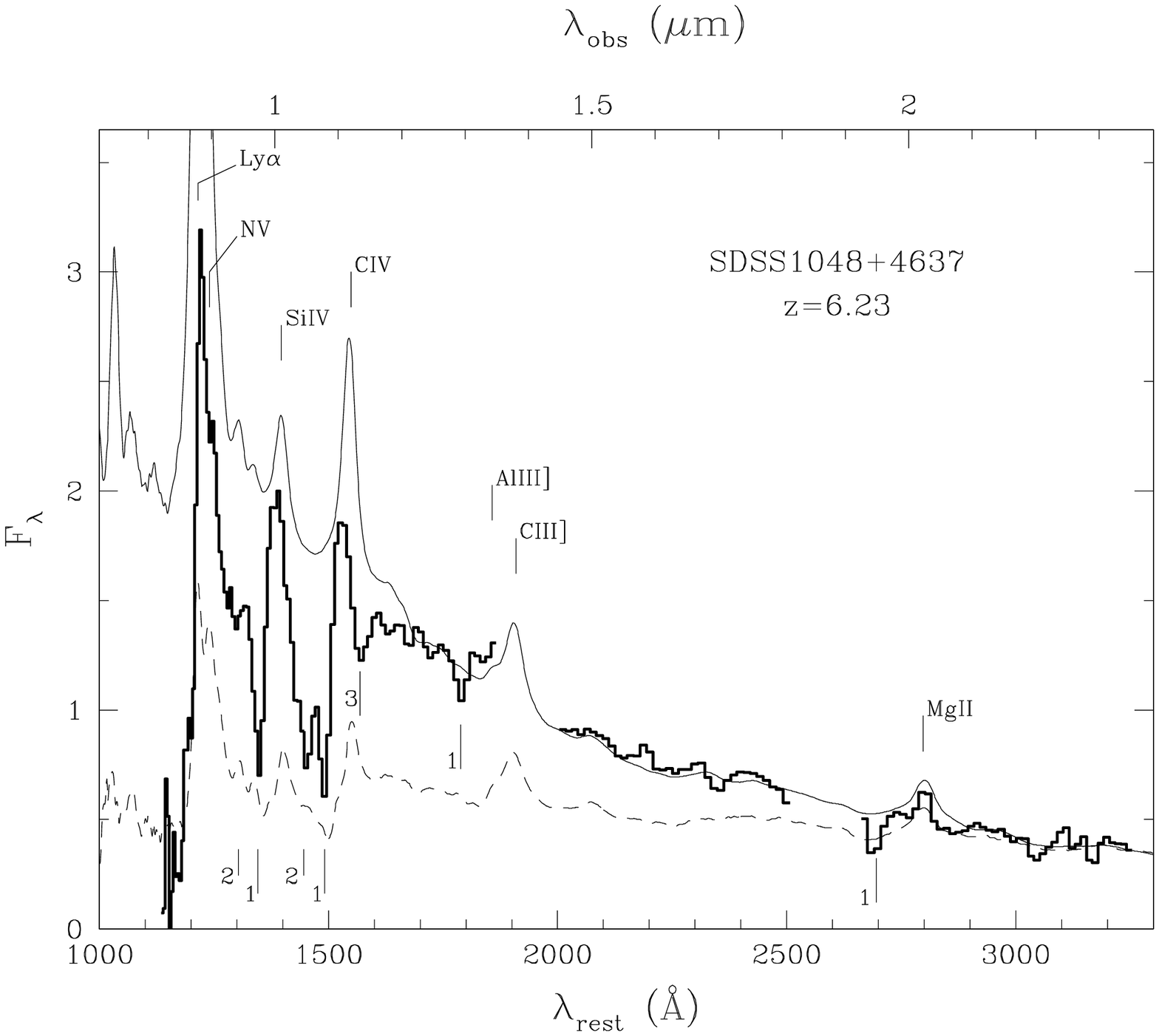}
\caption[]{{\it Left.} FeII/MgII observed in quasars as a function
of redshift. Solid symbols at high redshift are from our survey, while open
symbols at lower redshift are from \cite{dietrich03}
{\it Right.} The most distant LoBAL at z=6.23 found
in our sample. Note the extremely deep, blueshifted CIV absorption and the associated
absorptions of SIV, AlIII] and MgII. The thin solid line is the average spectrum
of lower redshift non-BALs, while the dashed line is the average spectrum
of lower redshift LoBAL from the SLOAN \cite{reichard03}. }
\end{figure}

We have investigated the metallicity and absorption of the most distant
quasars by means of the Near Infrared Camera Spectrometer
(NICS \cite{baffa01}) at the Telescopio Nazionale Galileo (TNG), with
its low resolution, high sensitivity IR spectroscopic
mode, which allows to obtain the full near-IR
spectrum from 0.8$\mu$m to 2.4$\mu$m in one shot.
Thanks to the large spectral coverage
and to the high sensitivity we could measure the intensity of the
UV FeII-bump (redshifted into the near-IR) relative to the MgII doublet (2798\AA)
in a sample of 22 quasars at $\rm 3<z<6.4$.
To a first approximation, the ratio FeII/MgII can be assumed as an
indicator of the Fe abundance relative to the $\alpha$ elements. This
is also an indicator of the age of the stellar population since $\alpha$
elements are predominantly produced by type II SN, while the production
of Fe is delayed by type Ia SNe \cite{matteucci01}\cite{romano02}.
Fig.3 shows the ratio FeII/MgII in quasars as a function of redshift,
indicating no decrease of the iron fraction up to z$\sim$6 (and actually
there is a marginal indication for an increase of Fe). This result has two
implications: 1) at the redshift of the most distant quasars known (z$\sim$6)
the circumnuclear medium was already highly enriched (in particular plenty
of Fe was available to provide obscuration in the X-rays); 2) the high ratio
of Fe/$\alpha$ suggests that the hosts of the most distant quasars were
formed at z$>$9, to allow the required time for the SN~Ia to enrich the gas.
Further details are given in \cite{maiolino03b}.

For eight quasars at 4.9$<$z$<$6.4 our spectra include the resonant
CIV line at 1549\AA .
Half of these quasars are characterized by deep, broad and blueshifted
absorption of CIV, i.e. they belong to the class of Broad Absorption Line
(BAL) quasars, which are associated with powerful outflows of dense gas.
Fig.3 shows the spectrum of the most distant of these BAL quasars,
characterized by a very deep absorption of CIV and also of other high and low
ionization lines (SiIV, AlIII, MgII), actually identifying this as
a Low ionization BAL (LoBAL). Although the sample is small,
the large fraction of BAL quasars, the depth and ionization state of
the absorption features suggest that these most distant quasars
are surrounded by a much larger amount of dense gas than lower redshift
quasars. As discussed in \cite{maiolino03c}, such a result may indicate
that the highest redshift quasars are characterized by extremely high
accretion rates and associated with the early formation of quasars and of their
host galaxies \cite{granato03}.

Another interesting finding is that all these distant BAL quasars
are bluer than lower redshift BALs (which are generally
reddened by dust). The dashed line of Fig.3 shows the average spectrum
of LoBALs at lower redshift from the SLOAN \cite{reichard03}, which is clearly
much redder than our highest redshift LoBAL (although at
$\lambda _{rest}< 1500$\AA \ there is a bending of the spectrum, partly
due to the prominent absorption lines, this issue is discussed
in \cite{maiolino03c}). The lack of significant
reddening in the most
distant BALs cannot be ascribed to lack of dust, since the presence of
large amounts of dust were inferred
by the submm/mm detections of these objects
\cite{bertoldi03}. Therefore, these results may indicate
a different extinction, and possibly reflect a different evolution
and formation mechanism of dust grains at z$>$5. This issue will be discussed
in \cite{maiolino03c}

{\it Acknowledgments.} The results presented in this paper were obtained
in collaboration with several people and more specifically:
S. Bianchi, T. B\"{o}ker, E. Colbert, A. Comastri, R. Gilli, F. Ghinassi,
G.L. Granato, Y. Juarez, A. Krabbe, F. Mannucci,
A. Marconi, G. Matt, R. Mujica, N. Nagar, E. Oliva,
M. Pedani, M. Salvati and L. Silva.

%

\end{document}